# Solar Energetic Particles: Spatial Extent and Implications of the H and He Abundances

**Donald V. Reames** (https://orcid.org/0000-0001-9048-822X)
IPST, University of Maryland, College Park, MD, USA

**Abstract** One of the earliest indicators of the importance of shock acceleration of solar energetic particles (SEPs) was the broad spatial extent of the "gradual" SEP events produced as the shock waves, driven by wide, fast coronal mass ejections (CMEs), expand across the Sun with cross-field transport mediated by the shocks. Contrasting "impulsive" SEP events, with characteristic enhancements of $^{3}$He and of heavy elements, are now associated with magnetic reconnection on open field lines in solar jets. However, large shock waves can also traverse pools of residual impulsive suprathermal ions and jets can produce fast CMEs that drive shock waves; in both cases shocks reaccelerate ions with the "impulsive" abundance signatures as well as coronal plasma. These more-complex events produce "excess protons" that identify this process, and recently, differences in the distribution of $^{4}$He abundances have also been found to depend upon the combination of seed population and acceleration mode. Extreme differences in the $^{4}$He abundances may reflect underlying differences in the abundances of the coronal regions being sampled by solar jets and, surprisingly, SEP events where shock waves sample two seed-particle populations seem to have about twice the $^{4}$He/O ratio of those with a single source.

# 1 Introduction

The evidence for solar energetic particles began with the largest events, where GeV protons caused nuclear cascades of particles through the Earth's atmosphere that were observed by ion chambers at ground level by Forbush (1946). The solar energetic particles (SEPs) in these ground-level events (GLEs) were seen above the continuous background produced in the same manner by the galactic cosmic rays (GCRs), and the peaks of the SEPs preceded broad decreases in the GCR intensity, called Forbush decreases, that we now understand to be caused by screening of the GCRs by magnetic fields of coronal mass ejections (CMEs) expanding outward from the Sun. Yet it would be decades before Kahler et al. (1984) showed a 96% correlation between SEPs and those same wide, fast CMEs that drove shock waves that actually accelerated the particles in these large "gradual" SEP events. A great controversy ensued before this important SEP source could be established in preference to an earlier presumed association with solar flares that were easier to see. This controversy regarding the "solar-flare myth" vs. the strong CME association is now widely documented (Gosling 1993, 1994; Reames 1995b, 1999, 2013, 2021b, 2021c; Reames et al. 1997; Zank et al. 2000; Kahler 2001; Cliver et al. 2004; Lee 2005; Cliver and Ling 2007; Rouillard et al. 2011, 2012; Gopalswamy et al. 2012; Lee et al. 2012; Desai and Giacalone 2016; Kouloumvakos et al. 2019).

The earliest measurements of element abundances accelerated in SEP events were made using large-area nuclear-emulsion detectors on sounding rockets (Fichtel and Guss 1961; Bertsch et al. 1969) and there were early attempts to compare the SEPs with abundances in the photosphere and corona (Biswas and Fichtel 1964; Webber 1975). Meyer (1985) compared the ratio of SEP and photospheric abundances and found two effects: (i) an overall dependence upon the first ionization potential (FIP) of the elements in all events and (ii) a dependence upon the magnetic rigidity or mass-to-charge ratio of the elements, $A/Q$, which varied from event to event (e.g. Reames 2018b). The FIP-effect comes from the corona itself and is a 3x enhancement of low-FIP (<10 eV) elements that are already ionized in the photosphere relative to high-FIP elements that begin their journey up to the corona as neutral atoms (e.g. Laming 2015; Reames 2018a; Laming et al. 2019). The $A/Q$ dependence was observed to be a power law (Breneman and Stone 1985) that also depended upon the source plasma temperature (Meyer 1985) through variation of the ionization states $Q$ with temperature.

When SEP abundance measurements on satellites became routine, some small SEP events had e.g. $^{3}$He/$^{4}$He = 1.52 ± 0.10 (Serlemitsos and Balasubrahmanyan 1975; Mason 2007), an enhancement by a factor of >3000 compared with the Sun or solar wind. Smaller ratios seen earlier were first attributed to nuclear fragmentation, as is seen in the GCRs, but lack of $^{2}$H or Li, Be, and B fragments from C, N, and O < 2 × 10$^{-4}$ (McGuire et al. 1979; Cook et al. 1984), showed strong evidence against fragmentation. These $^{3}$He-rich events involved a new wave-particle resonance phenomenon (e.g. Fisk 1978; Temerin and Roth 1992) and they were accompanied with steep power-law enhancements vs. $A/Q$ seen first up to Fe (e.g. Reames et al. 1994), and then as $(A/Q)^{3.6}$, on average, up to Pb when heavy-element measurements became available (Reames 2000; Mason et al. 2004, Reames and Ng 2004; Reames et al. 2014a). These $^{3}$He-rich or "impulsive" SEP events were associated with streaming solar electrons producing type III radio bursts (Reames et al. 1985; Reames and Stone 1986) in contrast with the type II bursts that were associated with shock waves and gradual SEP events, two SEP components noted long ago in a review of solar radio bursts by Wild et al. (1963).

Then Kahler et al. (2001) found that impulsive SEP events could also be associated with narrow CMEs and with solar jets, where magnetic reconnection on open magnetic field lines could allow easy escape of the SEPs without nuclear reactions. Steep power-law enhancements in *A/Q* were found theoretically in particle-in-cell simulations where ions are Fermi-accelerated as they reflect back and forth from the ends of the collapsing islands of magnetic reconnection (Drake et al. 2009). Impulsive SEP events can now be traced to their sources (Nitta et al. 2006; Wang et al. 2006; Ko et al. 2013; Reames et al. 2014a) and the SEPs are routinely associated with individual solar jets (Bučík et al. 2018a, 2018b, 2021). See the review by Bučík (2020).

While jets involve magnetic reconnection on open field lines, flares are driven by reconnection on closed field lines where the accelerated particles are also $^3$He-rich (Mandzhavidze et al. 1999; Murphy et al. 2016) and Fe-rich (Murphy et al. 1991), as determined from γ-ray-line measurements, but the ions, including the products of nuclear reaction that emit the γ-ray lines, like Li, Be, and B, cannot escape from flares. The hot bright flash in a flare is a direct result of energy containment. However, large solar events often involve reconnection on both open and closed field lines because type III bursts, produced on open fields in jets, often accompany large flares.

However, the clear distinction between impulsive and gradual SEP events had begun to fade. Not only did many jets have CMEs fast enough to drive shock waves (e.g. Kahler et al. 2001; Bučík 20 20), but also the large shocks of gradual events could reaccelerate seed populations from pools that had collected residual impulsive suprathermal ions (Reames 2022). Persistent pools have been widely observed and are often fed by many impulsive events too small to be distinguished individually (Richardson and Reames 1990; Desai et al. 2003; Wiedenbeck et al. 2008; Bučík et al. 2014, 2015; Chen et al. 2015). Reames (2020a) suggested four subtypes of SEP events leading to the differing abundance patterns:

(i) SEP1 events derived from magnetic reconnection in jets without shocks,

(ii) SEP2 events from jets with added acceleration by a local CME-driven shock,

(iii) SEP3 events where fast, wide shocks sample SEP1 pools and ambient corona,

(iv) SEP4 events where fast, wide shocks are dominated by ambient coronal ions.

In this article we first consider the spatial distribution of SEP events and particle transport from a modern perspective and then review the element abundance patterns that have led to defining the physical processes that lead to the four abundance patterns of SEP events. After showing the power-law abundance enhancements vs. *A/Q* and their temperature dependence, we then focus on the special relationships we have found of the elements H and He to the power-laws defined by the other elements, depending upon the presence or absence of shock waves and the nature of the seed population. Unsuperscripted He should be taken as $^4$He throughout this article.

Data presented are mainly from the *Low Energy Matrix Telescope* (LEMT) on the *Wind* spacecraft (von Rosenvinge et al. 1995), but also from the *Low Energy Telescope* (LET) on the *Solar TErrestrial RElations Observatory* (STEREO) ahead (A) and behind (B) spacecraft (Mewaldt et al. 2008; Luhmann et al. 2008). Data from these instruments are available at https://cdaweb.gsfc.nasa.gov/sp_phys/. Data on CME speeds were obtained from the *Large Angle and Spectrometric Coronagraph* (LASCO) on the *Solar and Heliospheric Observatory* (SOHO) at https://cdaw.gsfc.nasa.gov/CME_list/.

## 2 Spatial Distributions of SEPs

The earliest evidence favoring shock acceleration in large SEP events was their extensive spatial distributions. Since the particles are largely constrained to follow the Parker spiral magnetic field lines out from their source, how could they spread so far away from a point-source flare with only modest random walk of field lines (Jokipii and Parker 1969)?

Early observers did not have access to spacecraft surrounding the Sun to measure distributions in a single SEP event, but they were eventually able to examine hundreds of events from a variety of source longitudes on the Sun instead (Cane et al. 1988; Reames 1999). Today it is sometimes possible to view an event as shown in Fig. 1.

The *Wind* spacecraft is well connected to the shock of a 2175 km s$^{-1}$ CME near W21 on 23 January 2012 as shown in Fig. 1 and the intensities rise rapidly there in comparison with those seen at STEREO A on the far western flank of the shock which rise slowly as its magnetic connection improves during solar rotation, especially after the shock passage were the intensities at *Wind* and STEREO A eventually merge. While the field lines in the figure cannot show time variation and disruption as the shock passes, they do suggest that when the shock approaches 2 AU, shown by the dotted circle in Fig. 1, connection to STEREO A (green) approaches the shock nose while connection to *Wind* (blue) moves away from it.

Meanwhile STEREO B on the east flank shows an early SEP increase that stops growing, perhaps because the weak shock on this flank soon moves off its connecting field line (red). Note that the solar source would be behind the western limb as viewed from STEREO B.

Particle acceleration must occur at a sufficiently high altitude and low density that the ions do not immediately lose their energy to Coulomb collisions. Shocks begin to accelerate ions from the corona at a radius of 2 – 3 $R_S$ (Reames 2009a, b; Cliver et al. 2004); impulsive SEPs may begin at about 1.5 $R_S$ (DiFabio et al. 2008). The abundances of SEPs have a different FIP pattern from that of the solar wind (Mewaldt et al. 2002; Kahler and Reames 2003; Kahler et al. 2009; Reames 2018a; Laming et al. 2019), so SEPs are not just accelerated solar wind but are an independent sample of the active-region corona. Also, in situ, shock-accelerated ions are best correlated with ions upstream of the shock (Desai et al. 2003), suggesting that the shocks continue to reaccelerate the same general population of particles as they propagate outward. Shocks also expand laterally (e.g. Rouillard et al. 2012) and turbulence at shocks enables them to carry particles across fields and to mix and spread them as we will see in the next section, adding another dimension to the importance of shock waves.

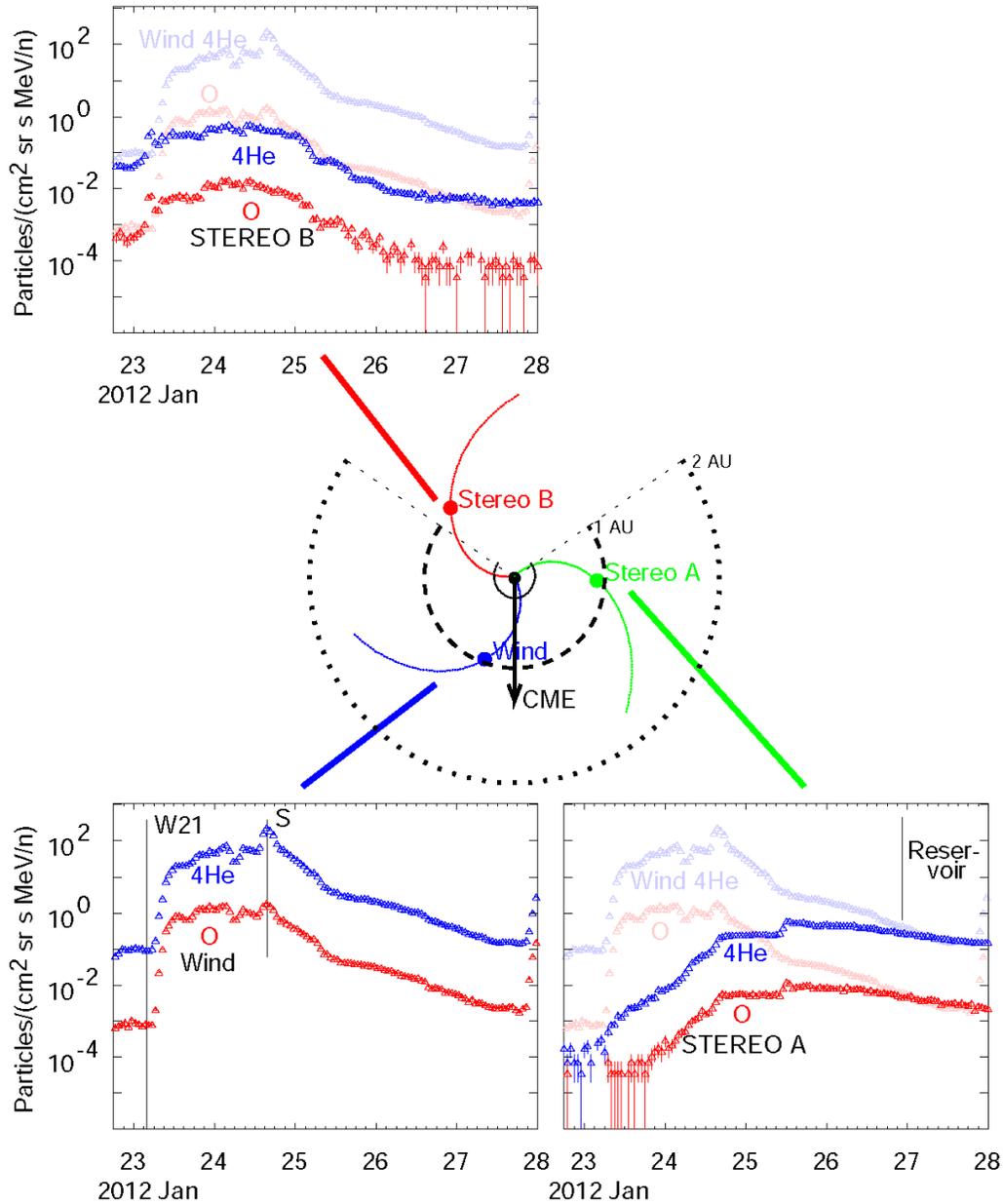

**Fig. 1** Intensities of $^4$He (*blue*) and O (*red*) near 4 MeV amu$^{-1}$ are shown vs. time for the 23 January 2012 SEP event as seen from the *Wind*, STEREO A, and STEREO B spacecraft, located around the Sun as shown in the central cartoon. The CME is directed downward along the arrow in the cartoon with a tentative opening angle shown for a spherically expanding shock. Faded time profiles from *Wind* are also shown in the STEREO panels for comparison. Intensities at *Wind* and STEREO A merge in a "reservoir" late in the event.

## 2.1 Cross-Field Transport: Shocks

To what extent do ions cross field lines? Curvature and gradient drifts are small. Pitch-angle scattering moves ions laterally only about a gyroradius at each scatter. However, while the particles generally follow field lines out from their source, there is certainly cross-field scattering in the immediate vicinity of the shock. In fact, a strong shock may approach the condition $\delta B/B \approx 1$ so the mean field direction is poorly defined. Waves must be sufficient to scatter and trap particles near the shock in order to accelerate them.

Figure 2 suggests the way turbulence at a shock can capture, mix, and distribute along the shock front, SEP samples from different longitudes. There will also be scattering and transport within the shock surface, but that mean free path may be short.

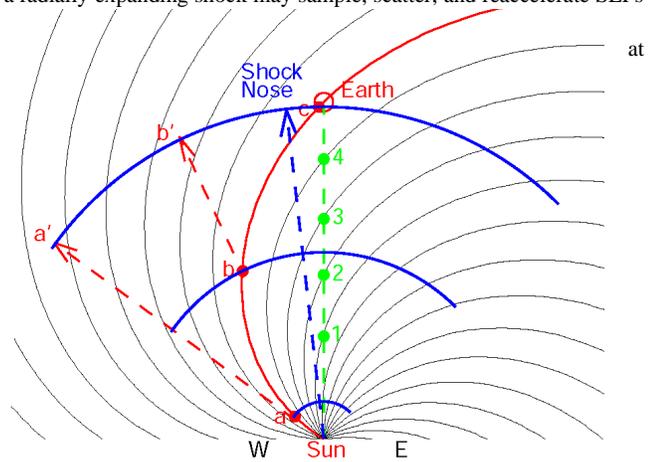

**Fig. 2** The drawing illustrates the way a given point on a radially expanding shock may sample, scatter, and reaccelerate SEPs from points on different field lines labeled 1, 2, 3, and 4, and may combine them into the intensity seen at Earth. Samples from the *red* field line at *a* can be carried by the shock radially out to *a'* and from *b* to *b'*, spreading the properties of the SEPs originally accelerated at *a* along the arc *a'*, *b'*, *c*. This assumes only pitch-angle scattering with no cross-field transport except in the immediate vicinity of shock itself and it ignores lateral shock spreading at its base.

Figure 2 assumes that particles previously accelerated on a given field line remain near that field line unless they are sampled and carried radially outward by turbulence at the shock, which mixes samples as it goes. Thus a fast shock moving out along the green corridor in Fig. 2 samples and mixes contributions from various field lines at points labeled *1*, *2*, *3*, and *4*, thus sampling nominally 55º of longitude, assuming a 400 km s$^{-1}$ pre-shock solar wind, by the time it arrives near Earth. Thus, the only true measure of initial acceleration at each longitude is that obtained early in an event. Later intensities are increasingly-large longitude averages – especially event-averaged intensities, which have blurred most of the longitude information.

Overall, the west flank of the shock will carry SEPs farther west across Parker-spiral field lines while, on the east flank, the Parker-spiral field itself will carry particles farther east as they follow the field outward; thus the longitude span of the SEPs increases with time, quite apart from spread of the shock itself. Mathematically, it is always possible to assume diffusion with adjustable coefficients for cross-field transport, but physically the process may be much more limited and yet more specific.

What about impulsive events? We might expect SEP1 events to be spread only by the smaller random walk of the field lines (Jokipii and Parker 1969), or by wandering flux tubes twisted within a CME, both of which we have ignored above. These events are generally found to be "scatter free" with scattering mean free paths of order 1 AU or more (Mason et al. 1989). However, SEP2 (and even SEP3) events may have spatially limited sources, but they also have shocks. If the turbulence at the shock is strong enough to reaccelerate ions, it is strong enough to carry ion samples radially. Specifically, it can carry the source ions from a single longitude, as the red curve in Fig. 2, from *a* to *a'* and from *b* to *b'*, spreading the ions from *a'* to *c* along 1 AU. Impulsive events with fairly wide distributions have been observed (Wiedenbeck et al. 2013).

Visible evidence of the shock transport across the Parker spiral that we are discussing is shown in Fig. 3 by the intensities at *Voyager 2*, poorly connected to the far western flank of the SEP event of 1 January 1978. No protons arrive early at *Voyager* at 1.95 AU along its poor initial magnetic connection to the shock, compared with the number arriving at the shock peak four days later. This is *not* "perpendicular diffusion." It is shock acceleration and transport. No particles are encountered initially, but they are soon transported across the Parker spiral to the *Voyager* field line by its connection to stronger, more-and-more central regions of the shock. As the magnetic connection from *Voyager* sweeps eastward along the face of the expanding shock with time, the proton intensity increases until the real culprit spreading those protons is suddenly seen when the peak at the shock itself passes the spacecraft. SEPs are quasi-trapped at the moving shock peak, historically called an "energetic-storm-particle" (ESP) event, while being reaccelerated. This shock is too weak to maintain 100-MeV protons so far from the Sun, but still significantly enhances those up to 22 – 27 MeV.

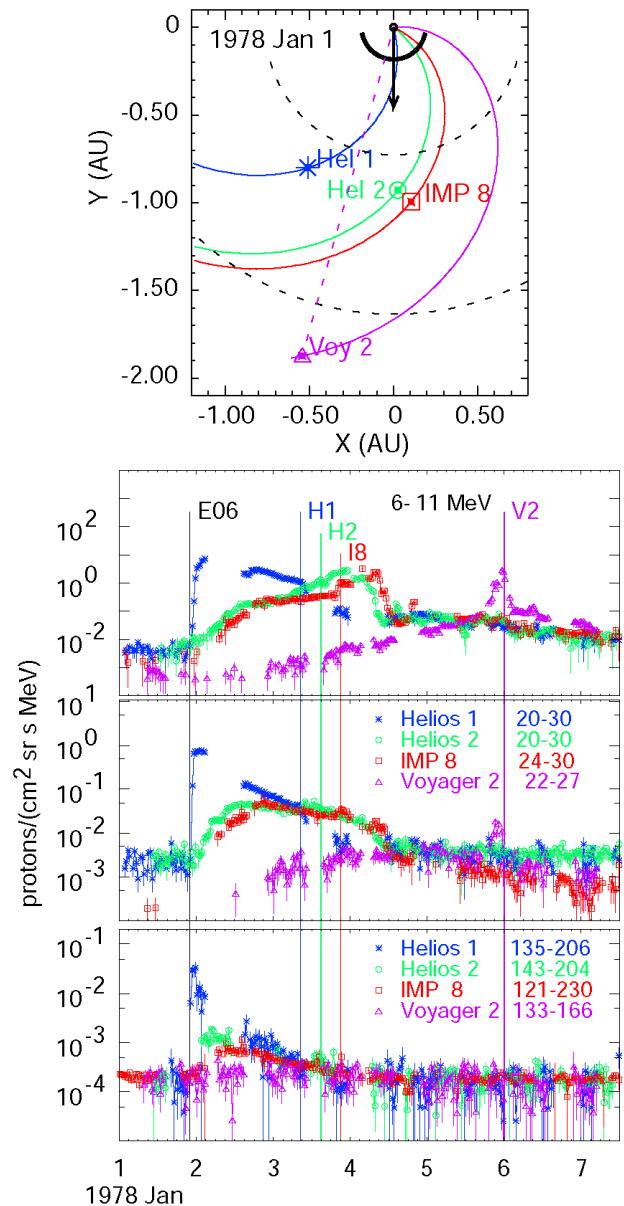

**Fig. 3** The upper panel shows the configuration of *Helios 1*, *Helios 2*, IMP 8, and *Voyager2* during the GLE on 1 January 1978 relative to the CME directed downward along the arrow. Colored initial field lines connect spacecraft to the Sun and dashed circular arcs suggest expansion of the CME-driven shock wave. Lower panels compare the intensities of protons observed at the indicated energies while vertical lines represent the flare-onset longitude E06 and shock-arrival times at each spacecraft (Reames et al. 2012). Protons are brought to the field line of *Voyager 2* by its improving connection to the expanding shock (or ESP event), *not* by "perpendicular diffusion." Below ~30 MeV, the shock peak at *Voyager* is nearly as intense as those seen earlier, but less well resolved, at *Helios 1*, *Helios 2*, and IMP 8. Unlike the other spacecraft, the field line from *Voyager* did not encounter the shock to receive particles initially.

## *2.2 Reservoirs*

Reservoirs were first observed by McKibben (1972) who noted equal intensities of 20 MeV protons over a wide longitude span on *Pioneer* spacecraft. Reservoirs were named by Roelof et al. (1992) who saw equal SEP intensities over radial distances from 1 to 2.5 AU late in large SEP events. Reservoirs were also seen using *Helios* (Reames et al. 1997; Reames 2013, 2021b) where identical proton spectra were seen at different spacecraft with intensities that decayed adiabatically as might occur for a magnetically trapped population of ions in a large volume which is expanding with time. Naively we might think that high-rigidity particles would scatter less, encounter the boundaries more often, and leak away from reservoirs faster, steepening the spectra with time, but this does not occur; spectra in reservoirs retain their shape and decrease adiabatically.

Figure 4 shows a way to exhibit a reservoir from a single spacecraft. Suppose we normalize the intensities of different species and different energies at a single time. In a reservoir, these intensities retain their normalization as all of them decrease uniformly with time. Other aspects of reservoirs are reviewed by Reames (2013); they were well explored by the *Helios* spacecraft which were more closely spaced than STEREO.

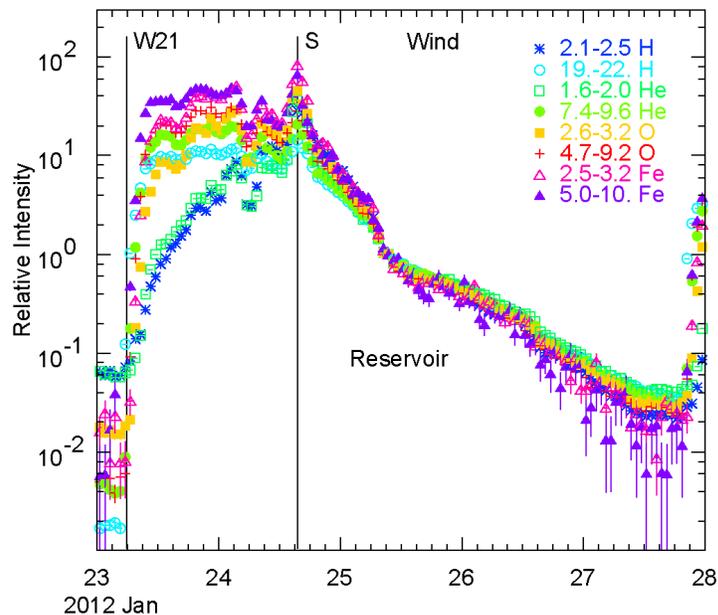

**Fig. 4** Intensities of various ions at various energies in MeV amu$^{-1}$ are normalized at one point in time during the reservoir of the 23 January 2012 event seen in Fig. 1, remaining similar throughout the reservoir. Ahead of the shock, the Fe, O, and higher-rigidity H and He propagate outward easily while strong scattering traps the low-rigidity H and He near the shock.

## 3 Element Abundances and Temperatures

The characteristic enhancement of heavy-element abundances in impulsive SEP events has provided a dependable means of identifying them. Figure 5a shows that separate impulsive and gradual peaks can be resolved in an unbiased sample of abundances for all 8-hour intervals with sufficiently high intensities during 19 years. The peak showing the impulsive enhancement of Fe/O is the basis for selection of individual impulsive SEP events (i.e. Fe/O/0.131 > 4). Abundances of $^3$He/$^4$He are highly variable and strongly energy dependent (Mason 2007) and show much poorer ability to resolve different event classes, since even large shock events can incorporate some $^3$He (e.g. Mason et al. 1999) and there are no two peaks as seen for Fe/O.

Figure 5b shows average abundances seen by LEMT in impulsive SEP events divided by corresponding reference abundances from averages over gradual SEP events (Reames et al. 2014a) as a function of the mass-to-charge ratio $A/Q$ determined by charge states at ≈3 MK. A similarly steep power law is also seen using time-of-flight measurements below 1 MeV amu$^{-1}$ by Mason et al. (2004).

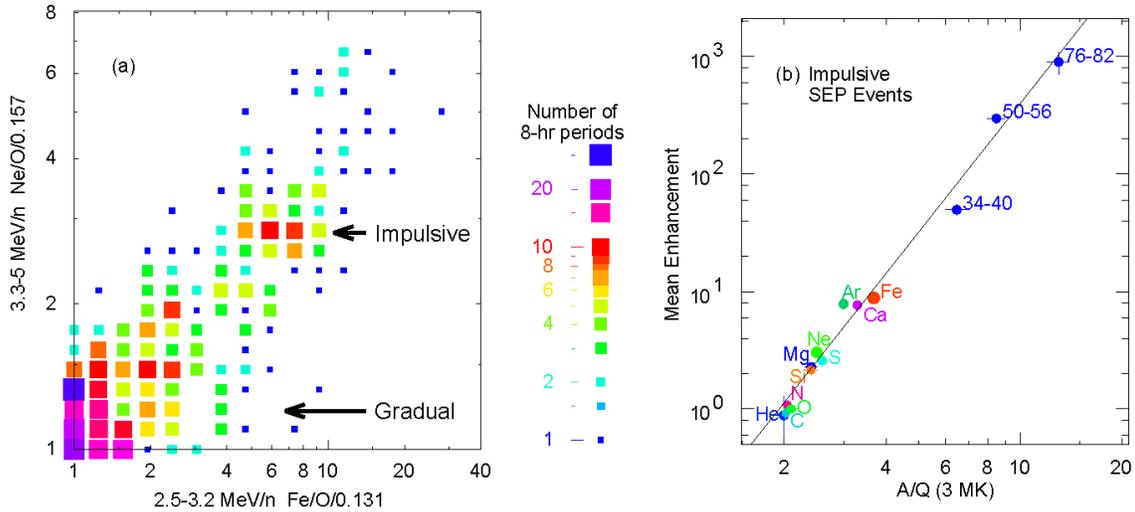

**Fig. 5** (**a**) shows a histogram of Ne/O vs. Fe/O normalized to the ratios in gradual SEP events and binned for all significant 8-hour intervals in a 19-year period. Panel (**b**) shows multi-event averaged abundances in impulsive SEP events divided by "reference corona" from averaged abundances in gradual SEP events near ≈3 MeV amu$^{-1}$ vs. ion $A/Q$ values at ≈3 MK (Reames et al. 2014a; Reames 2020a). H in SEP1 events continues the power law to $A/Q = 1$.

Once we accept the idea that the element abundances in individual events, divided by the corresponding coronal source abundances will have a power-law dependence upon $A/Q$ of the ions (e.g. Fig.5b; Reames et al 2014b, 2016a; Breneman and Stone 1985), resulting partly from a power-law dependence of scattering upon magnetic rigidity in large gradual events, we have a powerful new way to infer the $Q$ values, and hence the temperature of the source plasma. We have also assumed that using SEP abundances averaged over many gradual events provides a reference corona, which has been widely discussed by Reames (1995a, 2014, 2020a, 2021b) and compared with photospheric abundances to study the "FIP effect." In gradual events, differences in $A/Q$ during transport may cause some elements, e.g. Fe to scatter less than O, enhancing Fe/O early in events so that it becomes depleted later. However, such variations will tend to be averaged out with a large number of events viewed from many different times and longitudes.

The expected ionization states vs. temperature are determined from theory by Mazzotta et al. (1998) from H to Ni and by Post et al. (1977) at higher atomic numbers $Z$. Thus, as we choose each temperature, we can plot the abundance enhancements, relative to the coronal reference as seen for a small impulsive SEP event in Fig. 6b. Typical best-fit temperatures of 2.5 – 3.2 MK, determined for impulsive SEP events (Reames et al. 2014b), have been found to compare favorably with EUV temperatures determined for jet sources (Bučík et al. 2021).

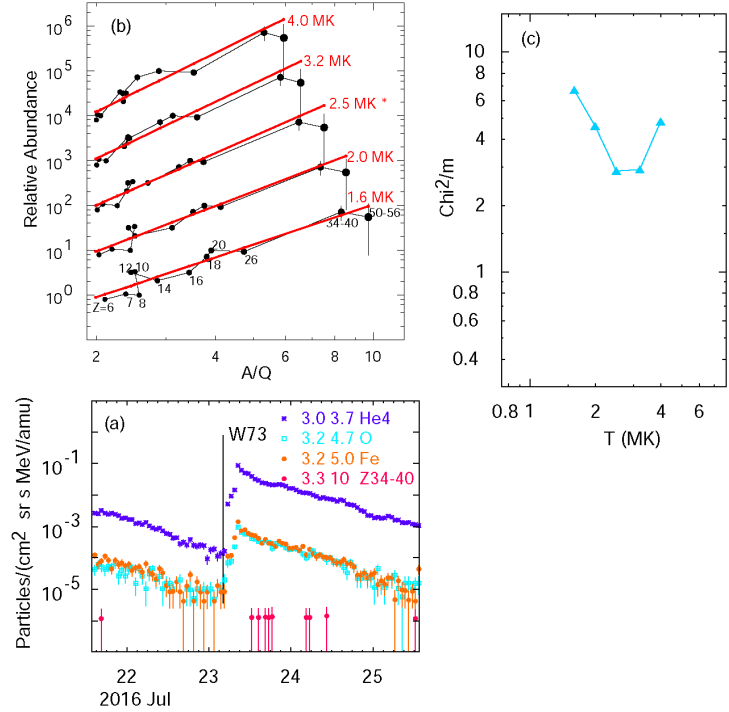

**Fig. 6** (**a**) shows the time variation of listed element intensities near 3 MeV amu$^{-1}$ for an impulsive SEP event beginning on 23 July 2016, (**b**) shows power-law fits of abundance enhancements of elements, noted by $Z$, during the event vs. $A/Q$ using $Q$ values (Mazzotta et al. 1998; Post et al. 1977) for the temperatures listed, spaced by factors of 10, and (**c**) shows $\chi^2/m$ values corresponding to each fit vs. temperature. A temperature of 2.5 MK is narrowly selected as the best fit with minimum $\chi^2/m$. The same abundance enhancements are plotted at each temperature, only the $A/Q$ values change (see Reames 2018b).

The ability to estimate source temperatures using element abundances greatly extends our characterization of SEP events, avoiding the effect of subsequent stripping that obscures the observed charge states of ions from impulsive sources (DiFabio et al. 2008), and providing temperatures found to be appropriate for jets (Bučík et al. 2021).

### *3.1 Proton Abundances and the Proton Excess*

What do we find when we extrapolate a power-law fit for $Z>2$ ions down to protons at $A/Q = 1$? Usually, the protons either fit fairly well or they miss by a wide margin.

For small impulsive SEP events we must collect the abundances from the whole event to get meaningful measurements of rarer elements (Fig. 6), but for more-intense gradual SEP events we can study time variations using daily or 8-hour averages. Figure 7 shows the best-fit power-law abundances for the 23 January 2012 event we studied in Figs. 1 and 4. Daily averages are shown for the less-intense, smaller-geometry STEREO instruments and 8-hour averages for *Wind*. The fits only include the $Z > 2$ elements, but are extrapolated down to H at $A/Q =1$. Best-fit temperatures at *Wind* and STEREO A give 1.6 ± 0.4 MK, while those at STEREO B are poorly determined (Reames 2017a). When the power laws are flat, i.e. there is no dependence upon $A/Q$, $Q$ values and temperature cannot be determined.

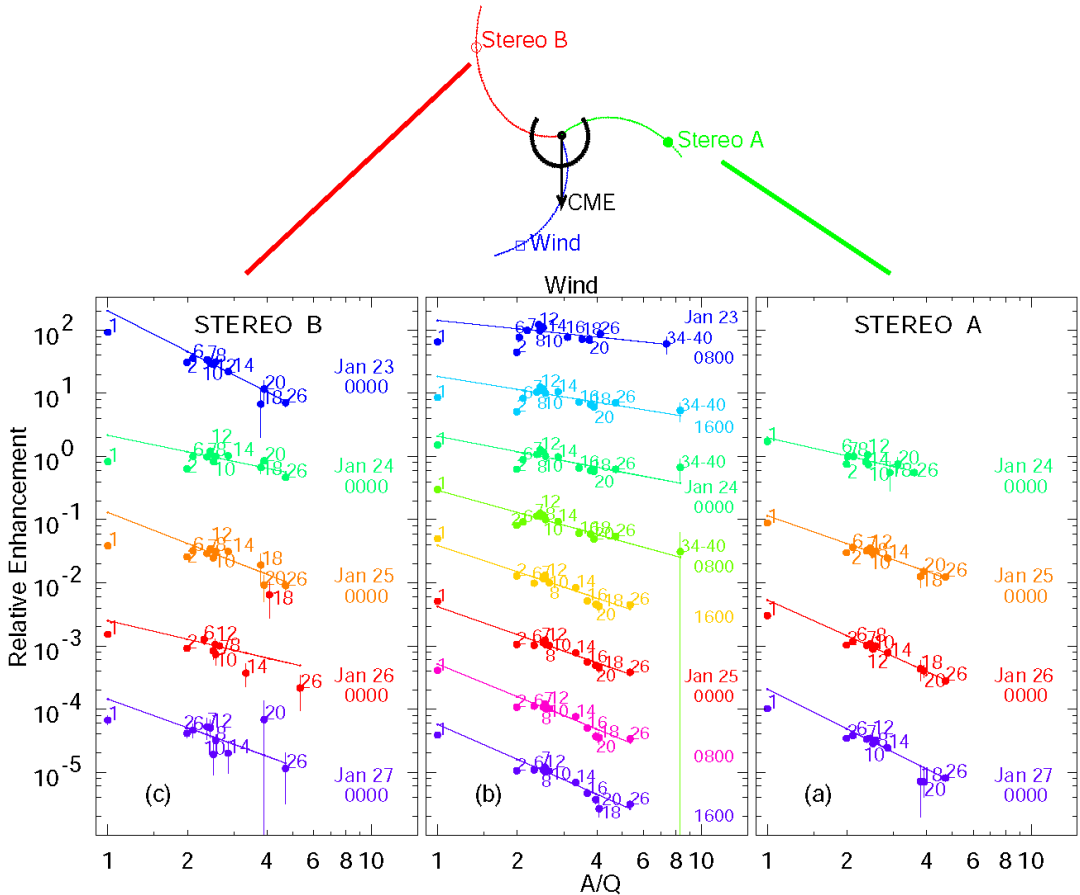

**Fig. 7** shows power-law fits to abundances of elements $Z > 2$, with elements noted by $Z$, beginning at the times listed for the STEREO and *Wind* spacecraft for the 23 January 2012 SEP event shown in Fig. 1. Fits of $Z >2$ ions intercept H at $A/Q = 1$ reasonably well, suggesting that all the elements come from a single source, i.e. ambient coronal material, in this SEP4 event.

　　　The most intense SEP events are SEP4 events where all element abundances are generally consistent with a single coronal seed population as in Fig. 7. However, Fig. 8 compares a SEP3 event, which has a substantial "proton excess", defined in the last time interval in Fig. 8c, with two large SEP4 events on the right, both GLEs, that lack proton excesses. We will see that the interpretation of the proton excess is that the dashed lines (approximately) in Fig. 8c are shock accelerated from ambient coronal seed ions, while the solid fit lines at higher $Z$ show residual impulsive suprathermal ions reaccelerated from collected pools that are widely observed (Richardson and Reames 1990; Desai et al. 2003; Wiedenbeck et al. 2008; Bučík et al. 2014, 2015; Chen et al. 2015). Thus the high-$Z$ enhancement in SEP3 events was already present in the seed population, while the high-$Z$ enhancements in the SEP4 events on the right in Fig. 8f are caused by scattering which allows higher-$Z$ elements to leak out to the observer preferentially while lower-$Z$ elements are trapped and suppressed ahead of the shock, also well studied (see Sect. 4; Ng et al. 2003) – H fits this pattern in the SEP4 events since it is from the same seed population. Both SEP3 and SEP4 events can be GLEs, since GLEs are controlled by energetic protons, not the pattern of high-$Z$ ions.

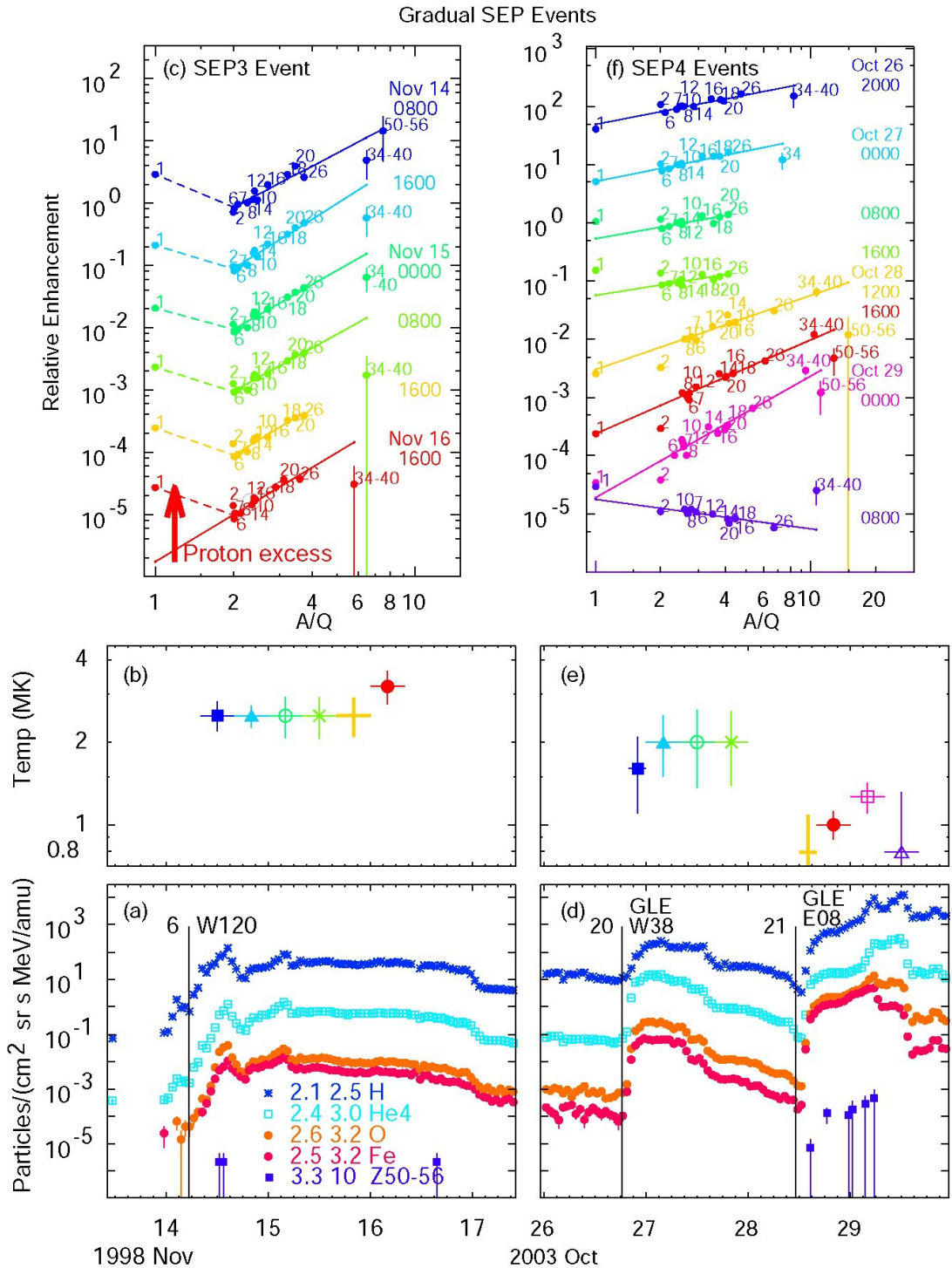

**Fig. 8** compares properties of a SEP3 event on the left with two SEP4 events on the right. Intensities are shown in panels (**a**) and (**d**), derived temperatures in (**b**) and (**e**), and power-laws in abundance enhancements vs. A/Q in (**c**) and (**f**) (Reames 2022). Event numbers in panels (**a**) and (**d**) refer to a list of gradual SEP events (Reames 2016, 2019c).

There are several examples where two successive SEP3 events, days apart, "double dip" into a single pool of suprathermal ions as it rotates across the face of the Sun (Reames 2022). Thus these pools of impulsive suprathermal ions clearly last many days. The ions may be weakly magnetically trapped in the high corona.

Impulsive SEP2 events also show proton excesses when CMEs from jets are fast enough to drive shocks that reaccelerate the SEP1 ions as well as ambient ions where protons dominate. Figure 9 shows an impulsive SEP2 event followed by a gradual SEP3 event. The CME speed was 824 km s$^{-1}$ in the first event and 1203 km s$^{-1}$ in the second. Note that the Sun rotates ~7º during the half day between the events, so the second event can easily sample the pool of impulsive suprathermal ions left by the first, as is seen quite often (e.g. Reames 2022).

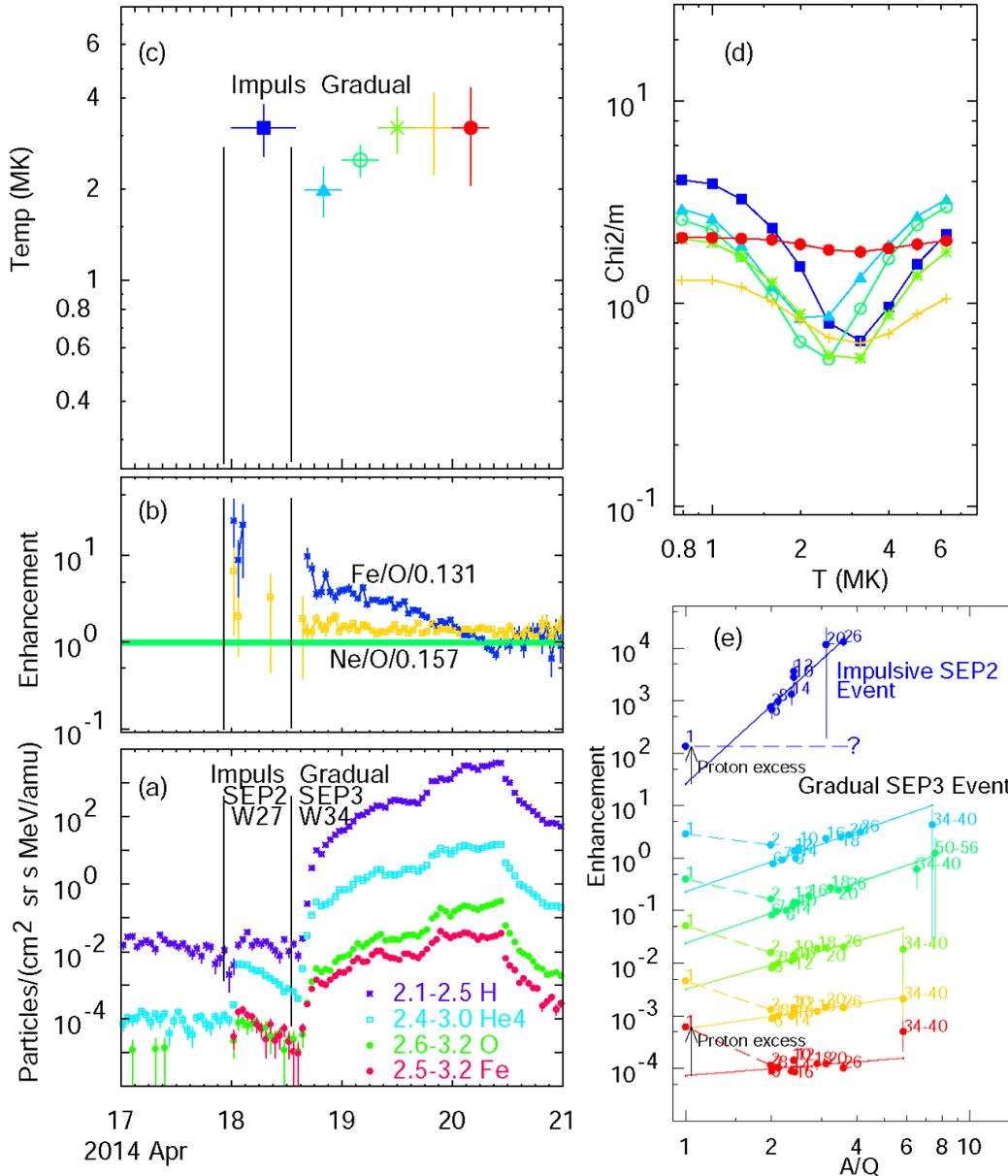

**Fig. 9** shows (**a**) time profiles of selected ions at listed MeV amu$^{-1}$ during impulsive and gradual SEP events, (**b**) Ne/O and Fe/O relative enhancements, (**c**) best-fit temperatures in selected intervals, (**d**) $\chi^2/m$ vs. temperature for each interval (color and symbols correspond with (**c**)) and (**e**) best power-law fits for $Z > 2$ ions, listed by $Z$ with fit extended to $A/Q = 1$, and offset x10 for each interval, with proton excess noted for the first and last interval (color coded with (**c**) and (**d**)). The '?' in (**e**) reflects the fact that ambient He, etc., is unlikely to increase with $A/Q$ as much as is allowed by the high He measurement.

The concept that shock acceleration of a two-component seed population might produce a proton excess is illustrated in Fig. 10 (Reames 2019b). The impulsive SEP1 suprathermal ions already have the built-in, $A/Q$-dependent enhancement of heavy elements while the ambient ions do not. Owing to the lower temperature of ambient corona of ~1 MK, vs. 2.5 MK for impulsive SEPs, C and O on the red line would be shifted to higher $A/Q$, i.e. lower $Q$, and suppressed somewhat by the downward slope.

It might be thought that the proton excess would be correlated with CME speed. However, increasing the shock speed would increase both components and the proton excess is basically the difference between the two.

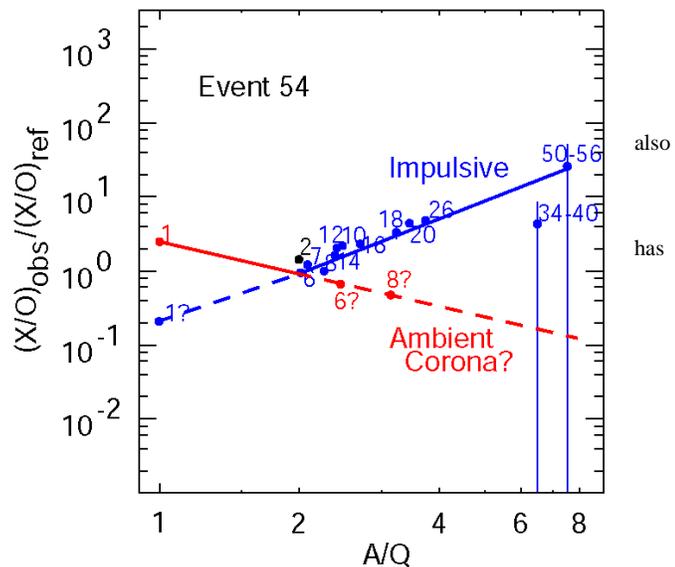

**Fig. 10** illustrates how impulsive-suprathermal (*blue*) and ambient coronal (*red*) seed populations might contribute when accelerated by a shock wave in either SEP2 or SEP3 events. For the ambient corona, heavier elements would be shifted to the right because of the lower temperature (lower $Q$s); this component could decline more steeply than shown, assuming both components must sum to contribute to He. The chosen example impulsive SEP2 event, Event 54 (Reames et al. 2014a; Reames 2019b), actually also has an associated CME speed of 952 km s$^{-1}$.

Thus the proton excess is a factor that measures of the ratio of the ambient and impulsive seed populations accelerated by a shock wave, i.e. it is the ambient-proton to impulsive-proton ratio. Some examples of its variations are shown in Fig. 11. Figure 11a shows an impulsive SEP2 event followed by two SEP3 events that are also both GLEs. The first two events have been examples in a classic comparison of impulsive and gradual events by Tylka et al. (2002). All three events in Fig. 11a come from a single region as it rotates across the face of the Sun and off the west limb (Reames 2022). The proton excess factors during these events are shown in Fig. 11c, and those for the third event show an unusual sustained decrease with time that might reflect a decreasing ambient/impulsive seed component spatial distribution on the east flank of the event as it rotates farther and farther behind the limb. While such conclusions should await a better understanding and modeling of the proton excess contributors, they do suggest the potential value of this kind of study.

The first event in Fig. 11d has too few heavy ions for definitive measurements, but seems to be an SEP3 event. The second event at E13 shows the highest proton excesses we have seen, which soon drop back near a value of one when particles from the large SEP4 event begin to arrive from the event centered at W11.

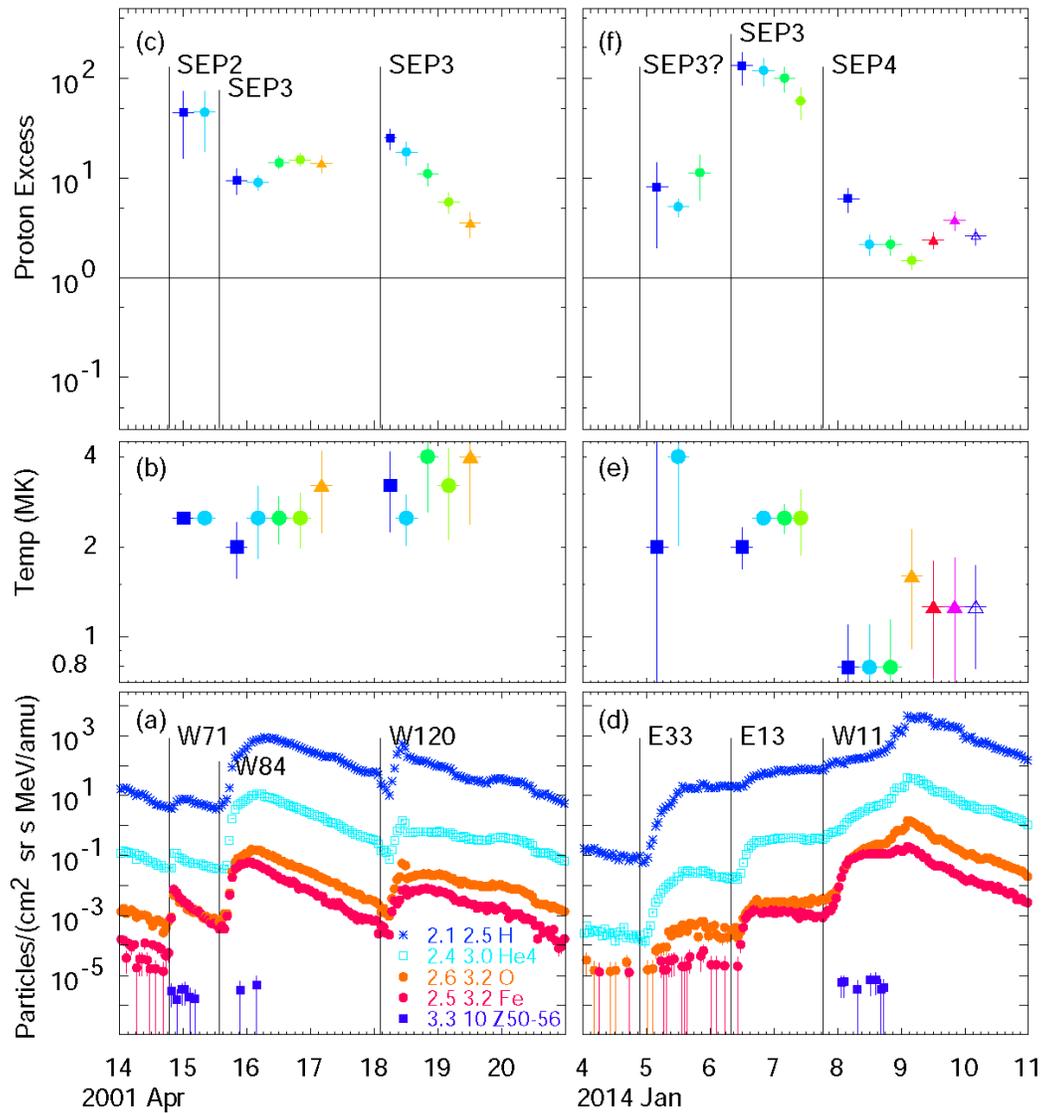

**Fig. 11**(**a**) and (**d**) show time profiles of selected ion intensities listed, (**b**) and (**e**) show derived source temperatures for selected intervals, and (**c**) and (**f**) show proton excess factors for the same intervals. Event onsets are flagged by source longitudes in the *lower* panels and by SEP class in the *upper* panels.

Note that when SEP2 or SEP3 events occur early in a sequence they may contribute their excess protons to the impulsive suprathermal ions in the pool. Thus, for subsequent events with shock waves that encounter that pool, their excess protons can come partly from the impulsive pool and partly from the ambient corona. This dilutes our information on the contribution of the two seed-particle sources in these late events that are shock-accelerated from a single pool. Thus, an early strong shock can reaccelerate SEP1 ions from a pool along with ambient ions that generate a proton excess. Suprathermal ions from this shock, along with their proton excess, are then added back to the pool creating a SEP2 pool that may be sampled again by any subsequent shock events. Effectively, a SEP1 pool can actually become a SEP2 pool.

Ions from impulsive SEP events are observed at 1 AU to be fully ionized up to S, presumably because they are stripped after acceleration (e.g. DiFabio et al. 2008).

However, any potential modification of the powers of *A/Q* by the shock in a SEP3 event is small compared with the steep power from the pool ions which may be averaged over many small SEP1 events, so the change in ionization may have little effect on the earlier enhancements correlated with the original *A/Q* values that persist.

### *3.2 He Abundances*

Abundances of He (unsuperscripted He means $^4$He), like those of H, show departures from the fit expected from the power-law from higher *Z*. Earlier studies that included He in the fits of gradual SEP events (Reames 2016a) caused high-temperature minima in $\chi^2$ that were erroneously interpreted as caused by stripping. This interpretation was completely wrong and was later (Reames 2017b) understood as source He abundance variations in gradual SEP events. However, other variations in impulsive SEP events are even larger, especially the occasional "He-poor" events (Reames et al. 2014a; Reames 2019a), so we no longer include He in fits. In Fig. 12 we show the impulsive event distribution in a space of He/O vs. power of *A/Q* for over 27 years of *Wind* data with individual plots for several selected extreme events shown around the border. Since the source temperatures for these events are all generally ≥ 2.5 MK, we assume that O is almost fully ionized in these events so that observed He/O is the source abundance, unaffected by any small *A/Q* changes. He/C would better meet this condition, but He/O is more familiar and any correction should be small for impulsive events. When these elements are fully ionized, they all have *A/Q* = 2, and we might expect their ratios to be unaffected by acceleration and transport; thus they would measure source abundances.

Events 1 and 125, with the lowest values of He/O in Fig. 12, characterize He-poor events. For Event 125 in Fig. 12c1 the observed intensity of Fe actually exceeds that of He. Neither of the He-poor events shown in detail have a proton excess, suggesting these are SEP1 events, without shocks.

The events in Fig. 12d are shown because Event 18 has the steepest power of *A/Q*, 6.4±0.9, of the 125 events. The event has a large proton excess but with a large error, 174±170, and an associated CME speed of only 293 km s$^{-1}$ so it may actually be a shockless SEP1 event.

Event 114 in Fig. 12e shows an unusually high He/O ratio. This event was also shown with its unusually high proton excess in the right-hand panels of Fig. 11. It is tempting to suggest that the shock-accelerated, ambient-coronal ions elevated both H and He in this event so He has received contributions from both impulsive and ambient seed populations, just as the protons did. Note that He is elevated in the SEP3 event in Fig.9e as well. This possibility is expanded below.

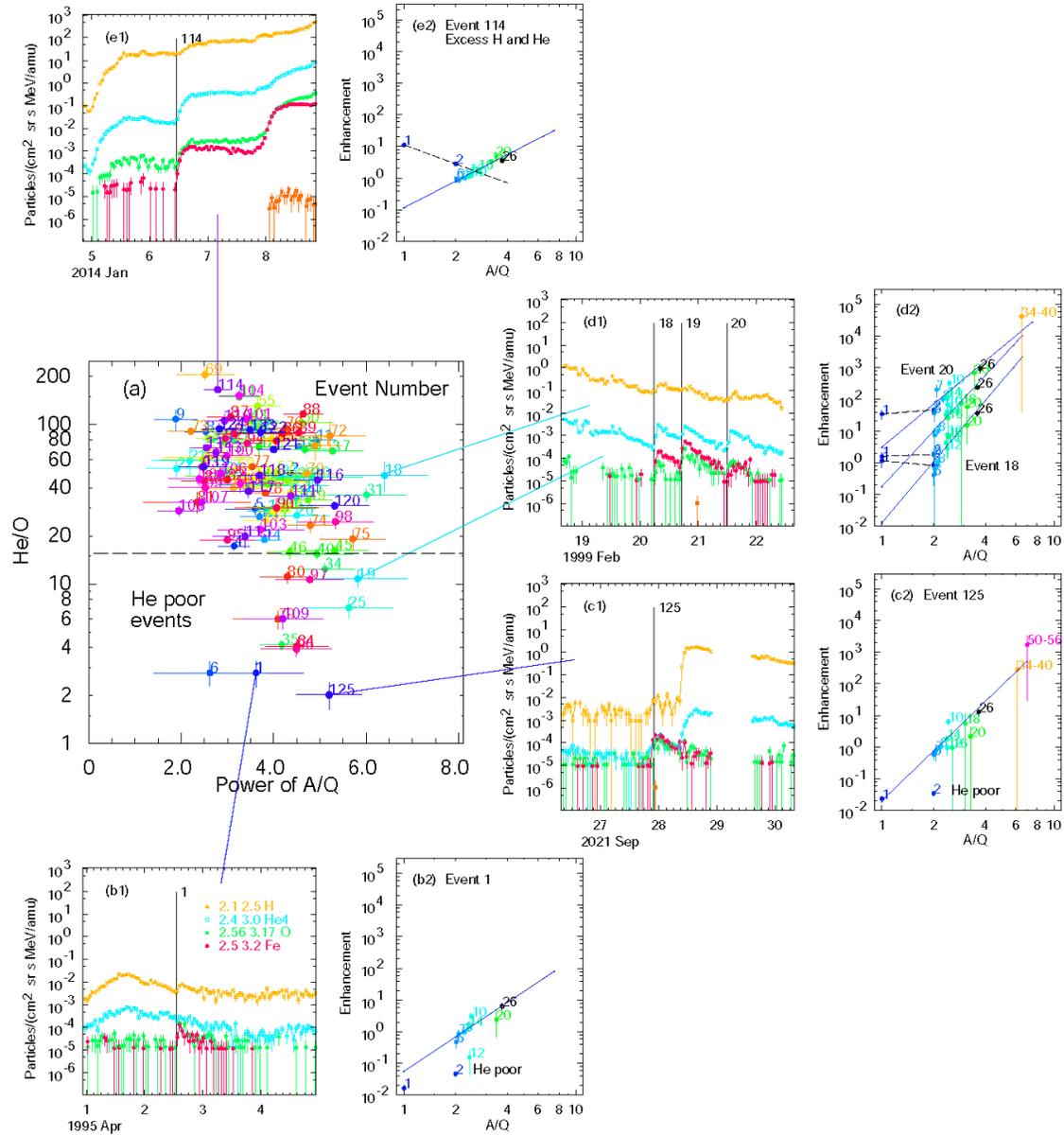

**Fig. 12** (**a**) shows He/O vs. power of *A/Q* for impulsive SEP events based upon the Fe/O criterion shown in Fig. 5a and listed by event number of Reames et al. (2014a. This event list has been extended by 14 events to 1 January 2022 using the same criteria.) *Panels* (**b**) – (**e**) show selected extreme events, (**b1**) intensities and (**b2**) power-law fits, etc.

The distribution of events along He/O is especially interesting since there are two large clusters of events, one near 45 and the other near 90.  If we divide the impulsive events into SEP1 and SEP2 events, where SEP2 events have a CME speed > 500 km s$^{-1}$ or, lacking a measured CME, have a proton excess factor > 5, we find the distributions of He/O shown in Fig. 13.  It is also true that among gradual SEP events the higher-temperature, impulsively-seeded SEP3 events tend to have source He/O near 90 while the ambient-seeded SEP4 events have lower He/O (see Figs. 6 and 12 in Reames 2018b). These distributions of He/O for SEP3 and SEP4 events are included in Fig.13.

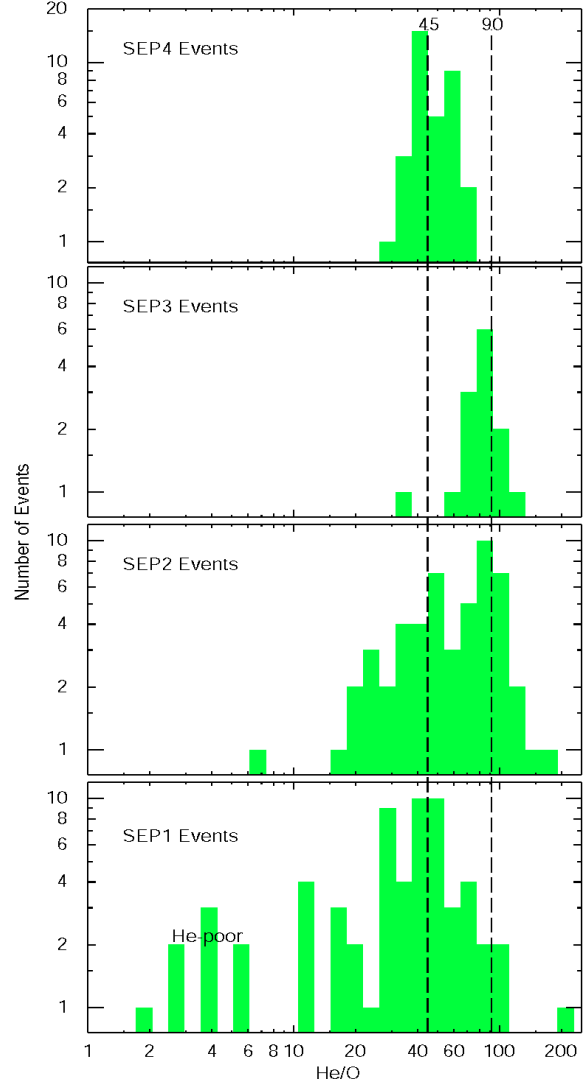

**Fig. 13** Histograms compare the distributions of source He/O values for impulsive SEP1 and SEP2 events and for gradual SEP3 and SEP4 events. Do the shocks in SEP2 and SEP3 events accelerate He from both seed populations while SEP1 and SEP4 events sample only one?

A common explanation for reduced He generally is that its uniquely high value of FIP = 24.6 eV causes depletion because it is slower to ionize as it crosses the chromosphere (e.g. Laming et al. 2009), slower even than other high-FIP elements. The small He-poor events do show some evidence of small spectral differences of ions (Reames 2019a), although not enough to explain the extent or the nature of the He suppression. Shock acceleration in SEP2 events might regularize any spectral variations. For reference, in the slow solar wind He/O ≈ 90 ± 30 and in the fast wind He/O ≈ 75 ± 20 (Bochsler 2009).

However, there is an alternate view of the He abundance that explains the peaks in He/O. We have normalized He/O enhancements for the impulsive events assuming a coronal value of $(He/O)_{cor} = 57$, the mean value derived from the gradual events. Suppose we changed the assumed coronal reference abundance to $(He/O)_{cor} = 45$ instead. Then the SEP1 and SEP4 events, with a single coronal seed population would tend to reflect this coronal reference abundance of 45, since this value is chosen to explain them, while all the SEP3 events and about half of the SEP2 events have twice as much He, presumably about half from each contributing seed population, impulsive and coronal. If we change the coronal abundance of He as suggested, the He points would move up 27% on all of the plots of enhancement vs. *A/Q*. (Note that would improve the He fit in Fig. 7, for example)

What is wrong with this alternate He abundance? (i) There is no known reason that the two seed components should provide comparable He contributions to produce a nice peak at He/O = 90. (ii) There is no evidence for a two-component correlated enhancement in other ions, e.g. C, etc., although, perhaps, none is expected. (iii) There is no other reason (He/O)$_{cor}$ should be so low for SEP sources; (He/O)$_{cor}$ = 90 is in much better agreement with theory (Laming et al. 2019; Reames 2020a) and with the solar wind. However, He/O = 45 is observed clearly in many events and must be explained in any case if there are really two peaks in the He/O distribution instead of a single broad distribution.

We can further investigate the distribution of He/O in impulsive events by comparing proton excess vs. He/O in Fig. 14. Unfortunately, proton intensities are obscured by background in many impulsive SEP events so we have only 75 events in the lower panel of Fig. 14, out of a total of 125 impulsive SEP events, so not all of the impulsive events in Fig. 13 can be studied. Nevertheless, the impulsive event correlation coefficient for proton excess vs. He/O is 0.57, suggesting that the H and the He indeed may sometimes both come from the same seed populations so they vary together. For the SEP2 events alone, the correlation coefficient remains at 0.44. SEP1 and SEP2 events have been distinguished either by the proton excess or by CME speed; the latter is shown in the upper panel of Fig 14 for those 49 events with both CMEs and measurable protons.

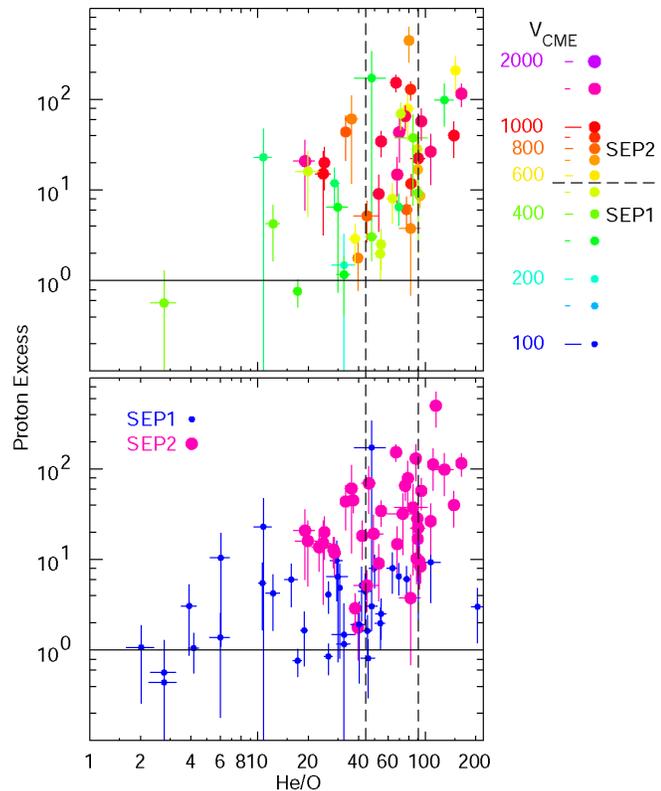

**Fig. 14** Shows proton excess vs. He/O for impulsive SEP events in both panels, with SEP1 and SEP2 events distinguished in the *lower* panel and CME speed shown by point color and size in the *upper*.

It has been found that the coronal FIP pattern in SEP events has initially evolved on closed magnetic loops around active regions, in contrast with the open solar wind (Reames 2018a, 2020a, 2021b; Laming et al. 2019). These comparisons assumed He/O ≈ 90, but the suppression of He because of its high FIP = 24.6 eV is well documented (e.g. Laming et al. 2009) and He is not critical to the differences between SEPs and the solar wind, which is based primarily upon different abundances of C, S, and P (Reames 2018a). Figures 13 and 14 show an especially wide spread of He/O in the SEP1 events. Perhaps solar jets occur in small local regions where the fractionation of He is incomplete, often He/O < 40, occasionally He/O < 10 and rarely He/O < 2. In this picture, the He-poor events are a consequence of incomplete FIP fractionation in some local coronal regions occasionally sampled by solar jets. All of the shock-driven events, SEP2, SEP3, and SEP4, which show more He, are intrinsically much more energetic than SEP1 events and likely involve ions sampled from much larger volumes of plasma so local variations tend to average out, especially for the SEP3 and SEP4 events.

In the SEP2 events, the shock accelerates ions from the local ambient corona as well as reaccelerating SEP1 ions. The ambient H easily dominates the highly-suppressed SEP1 H, but the He enhancement depends upon the relative contributions and power-law slopes of both seed components (see Fig. 10). In some events the ambient He seems to add little, in others it doubles the average SEP1 He, in Fig. 12e somewhat more, but then there seem to be a dearth of events from there to the SEP4 events where the SEP1 component is gone entirely and the ambient corona dominates all species.

This explanation of the He abundance, based upon $(He/O)_{cor}$ = 45 sampled by the SEPs, is tentative, and may turn out to be incorrect. Nevertheless, it is interesting and the peaks in He/O seem to correspond with the four previously-defined abundance categories of SEPs, i.e. those classes with shocks and two seed components have up to about twice the coronal He of those with a single seed.

## 3.3 $^3$He-rich Events

While $^3$He/$^4$He varies widely in impulsive SEP events, and even has a strong and variable energy dependence (Mason 2007), we show, for completeness, the location of the higher $^3$He-rich events among our sample of events. Figure 15 shows the distribution of events with $^3$He/$^4$He ≥ 0.1 in the space of $^4$He spectral index vs. power of *A/Q*. The events with high $^3$He/$^4$He tend to be smaller events with steeper energy spectra. Event 25 with $^3$He/$^4$He = 5.4±0.3 is also somewhat $^4$He-poor, but $^4$He for this event would be nearer normal if we let $(He/O)_{cor}$ = 45. H from the Event 25 is obscured by background.

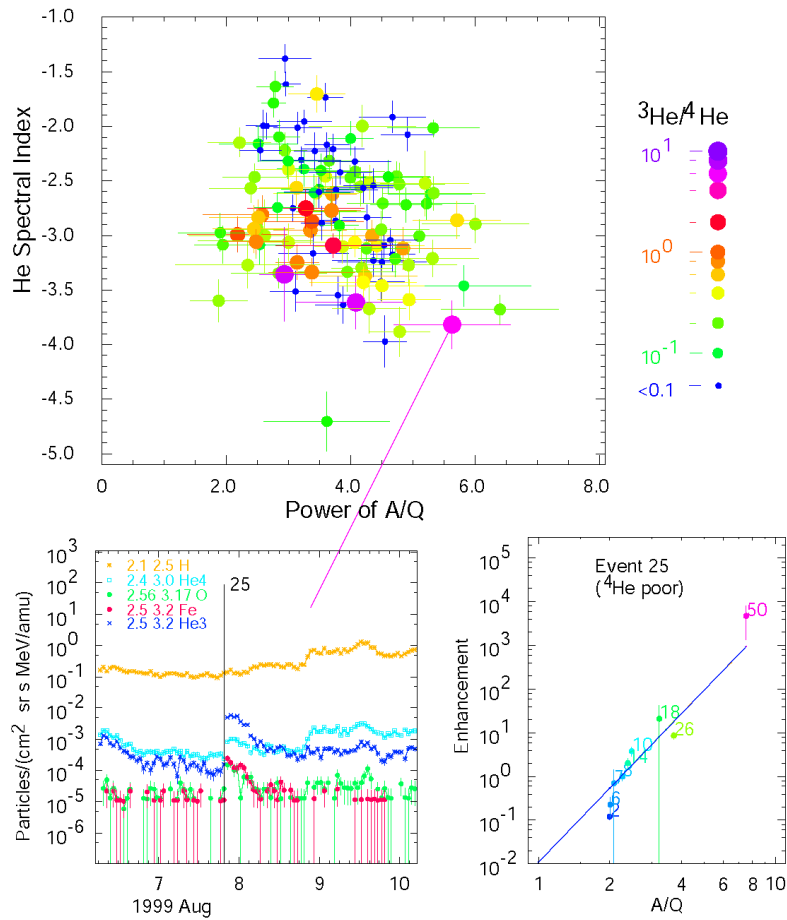

**Fig. 15** The *upper panel* shows symbol color and size for events with values of $^3$He/$^4$He >0.1 located in a plot of $^4$He spectral index vs. power of *A/Q*. The *lower panels* show intensities (*left*) and fitted power of *A/Q* (*right*), with points noted by *Z*, for an event with very high $^3$He/$^4$He.

## 4 Acceleration vs. Transport

An important question about SEP events is: which observed properties are caused by acceleration and which by transport (Reames 2020b)? Mason et al. (1989) found that impulsive events tend to propagate scatter free, meaning a scattering mean-free-path > 1 AU. They also found that scatter-free conditions sampled by a small $^3$He-rich event, exist during the declining phase of a large SEP event. There are also small gradual SEP4 events with minimal scattering, with the particles streaming outward for a day or more (Reames 2020b; Reames et al. 2001). However, there is a large range of events where transport seems to matter little, for example, the 23 January 2012 event shown in Figures 1, 4, and 7 shows very little change in abundances and energy spectra (not shown) with time or over 222º of solar longitude. Thus, wide spatial differences cause little change. Only a few low-rigidity ions, like 2 MeV H or He, show evidence of trapping upstream of the source shock in Fig. 4. Events of this character are quite common. The SEP3 events also show abundances that are unusually constant with time (Reames 2020a, 2021c, 2021d, 2022), while the SEP4 events show a relationship between energy spectral indices and powers of *A/Q* (Reames 2022).

Only the largest SEP4 events show strong transport effects in our energy range, where low-rigidity ions are especially trapped driving power-law abundances with positive slopes that persist until the shock passes, as in Fig. 8f (Reames 2020b). These are the events where the early streaming protons amplify Alfvén waves that scatter and trap ions, flattening the low-energy spectra at the "streaming limit" (Reames 1990; Ng et al. 1999, 2003, 2012; Reames and Ng 1998, 2010). These events can also reflect the presence of impulsive suprathermal ions (Tylka et al. 2005; Tylka and Lee 2006; Sandroos and Vainio 2007) but they do not dominate the event.

Wave-particle interactions leave many questions. If streaming ions generate waves, when are those waves damped? Are wave modes that are amplified upstream of shocks damped downstream? What is the ambient interplanetary wave intensity; is it the wave intensity sampled by scatter-free $^3$He-rich events? SEP diffusion theories often treat diffusion coefficients as adjustable parameters that are constant is space and time. This is surely a poor approximation in large SEP events (Ng et al. 2003) and perhaps in most events.

## 5 Perspective and Summary

Our treatment of the proton excess is really backward from the normal perspective of the selection of ions from the seed populations by the shock. Historically we measured and fit the rarer $Z > 2$ ions first, then we asked about protons. Logically, we should first consider that the shock independently accelerates the proton-dominated ambient coronal ions and, as adequate waves are established, mainly by protons, the shock reaccelerates any available impulsive ions. While the *relative* abundances of H may seem suppressed in some of our comparisons, protons completely dominate the *absolute* abundances, and far exceed all other ions in all of the SEP events. In the resulting SEPs, we should begin with these dominant protons, and then the relative abundances of impulsive ions may "stick up" at high $A/Q$ above the ambient ions. They stick up because these high-$A/Q$ ions were already enhanced in the seed population – they carry this signature from their SEP1 origin (Fig. 5b), but they contribute relatively few protons. They also carry some $^3$He enhancement (e.g. Mason et al. 1999) and they may also contribute to SEP4 events where they are not dominant. Most protons, and possibly some He, come from the same source in SEP2, SEP3, and SEP4 events; they are shock accelerated from the ambient plasma. The term "proton excess" comes from the historic, high-$Z$, SEP1 perspective; actually, the protons are quite normal and there may or may not be a "high-$Z$ excess". SEP4 events lack a "high-$Z$ excess"

A summary of the status of our understanding of SEP abundances is as follow:

1) Element abundances of the average coronal source of SEPs have been established over many years with measures of the "FIP effect" relative to photospheric abundances (e.g. Reames 1985, 2014, 2021a). This has served as a necessary basis for study of different systematic enhancements in individual SEP event which tend to behave as powers of $A/Q$.

2) The primary acceleration source for the impulsive or $^3$He-rich events has been established as magnetic reconnection on open field lines in solar jets. This acceleration produces steep enhancements of heavy elements, typically $(A/Q)^3$ (Fig. 5b), protons being suppressed by this same power. These are the pure SEP1 events. Derived ionization states, $Q$, suggest source temperatures of ~2.5 MK, values also found recently from EUV studies of SEP source jets (Bučík et al. 2021). $^3$He/$^4$He is enhanced by independent wave-particle resonance.

3) The dominant process in most large gradual SEP4 events is acceleration of ambient coronal ions by wide, fast CMEs. Abundances are sometimes enhanced as modest powers of *A/Q* but often qualify as "proton rich," *declining* as powers of *A/Q* (e.g. Fig. 7). Sources are nearer ambient coronal temperatures with values 0.8 – 1.8 MK.

4) Fast shocks can always find seed particles in the ambient corona, but often also encounter, and even favor, residual impulsive suprathermal ions, either in jets (SEP2 events) or in pools accumulated from many small jets in active regions (SEP3 events). Combining proton-rich ambient seed ions and proton-poor impulsive seeds produces dual-sloped enhancements with H, and occasionally $^4$He, dominated from the ambient seeds while *Z* > 2 ions are dominated from the impulsive seeds (as in Fig 10).

5) The fits to *Z* > 2 abundance enhancements vs. *A/Q* have been extremely important, not only in their own right – determining temperatures, etc. – but because they highlight departures of H and He, which have led to important new physics. For example, the H abundance can distinguish jets with and without shock reacceleration, (SEP2 vs. SEP1).

CME-driven shock waves are an important factor for SEP events in many ways. Of course, wide, fast CMEs drive shock waves that accelerate gradual SEP events and expand across a broad front, but also, in the presence of Parker spiral field lines, they mix and spread these SEPs in solar longitude. When shocks traverse and accelerate mostly ambient coronal plasma they produce SEP4 events. The high-*Z* abundances can increase or decrease with *A/Q*, partly from transport, allowing us to estimate a source temperature from the best-fit power law, but often the abundances are just coronal (i.e. flat) and the temperature is indeterminate. The source seed particles for the SEP3 events are the widely-observed, persistent, accumulated pools of impulsive suprathermal ions from multiple jets (Desai et al. 2003; Wiedenbeck et al. 2008; Bučík et al. 2014, 2015; Chen et al. 2015), they can produce SEP3 events dominated by spikes of impulsive high-*Z* ions with their built-in steep power-law increase with *A/Q*, but with excess H and He relative to the steep power law of impulsive ions.

Magnetic reconnection accelerates ions with steeply increasing power-law enhancements vs. source *A/Q*. Reconnection on closed field lines produces solar flares while reconnection on open field lines in jets allows SEP1 events to escape into space, thus we can observe them. In SEP2 events, jets with sufficiently fast CMEs produce shock waves that accelerate ions from the local plasma, especially seen in protons, and reaccelerate the SEP1 ions, producing the characteristic abundance enhancement at high *Z* that is a SEP1 signature.

SEP2 and SEP3 events both have excess protons relative to those seen for the reaccelerated SEP1 impulsive component alone. The protons are derived from shock acceleration of the ambient coronal plasma. There may also be similarly-derived He. SEP3 events always have He/O ≈ 90 and SEP2 events frequently do also. SEP1 events and SEP4 events are mostly near He/O ≈ 45.  However, about 10% of SEP1 events are He-poor with He/O < 10.  This behavior is not well understood but suggests that small jets may occur in limited spatial regions of the corona that have these diminished He/O ratios, perhaps where high-FIP He has been incompletely fractionated.  There is an interesting possibility that the average coronal He/O abundance accessible to SEPs is 45 as seen in SEP4 events and in most SEP1 events, while shocks in the SEP3 and most SEP2 events can get comparable He contributions from both of the two available seed populations, SEP1 ions and ambient plasma.  Thus the events with He/O ≈ 90 get both He and H from the ambient plasma; they manage to get twice the He, from the two seed populations, but cannot get twice the O.  Previously, we started by assuming a coronal value for He/O of 90 and asked how it could get suppressed in SEPs; this was not very successful.  Now we start with a coronal value of 45 and ask how it gets increased in SEPs; this seems better, but we have yet to explain this initial value.

       The four classes of SEP events incorporate much of the new information we have acquired on the physical processes of SEP acceleration involving jets, magnetic reconnection, CMEs, and shocks and the complexity of new and reprocessed seed populations for the latter.  Proton abundances were already an important factor in this story and we now find that He abundances depend significantly upon these classes (Fig. 13) for reasons we have only begun to assess.

## 6 A Few Questions

We suggest a few questions for future study:

1) In impulsive, SEP1, events, what is the relationship, if any, between the magnetic reconnection that enhances heavy elements and the wave-particle resonances that enhance $^3$He/$^4$He?

2) In the SEP4 events, what drives the correlation between power-law spectral indices and power-law abundance enhancements?  Shock compression ratios determine spectral indices; can they also determine relative sampling of H, O, Fe, and Pb?  Doesn't the selection of ions relate to the shock "injection problem" (Zank et al. 2001)?

3) Do the jets that produce the extremely low $^4$He/O, i.e. He-poor events, share some distinctive regions of the corona where high-FIP He is scarce?  Can we probe spatial variations in coronal abundances with SEP1 events?

4) Is there any non-SEP evidence for He/O ≈ 45 in the active-region corona?

5) Is it possible to build a physics-based model with a self-consistent theory of shock acceleration with time-dependent evolution of both particles and Alfvén waves and wave-particle interactions that predicts the spectra and abundances of all SEP particle species based upon known coronal abundances? A successful theory of this depth could inspire great confidence in both understanding and forecasting SEP events.

6) Is it possible to model ion acceleration during magnetic reconnection so we can better understand what factors control both the abundances and the spectra?

7) Diffusion theory is appropriate for particle pitch-angle scattering in parallel propagation, but is entirely inappropriate for cross-field transport dominated by shocks expanding across Parker spiral fields (e.g. Fig. 3). Can we remove "perpendicular diffusion" and substitute realistic shock transport in models of spatial evolution of SEPs?

**Acknowledgement** The author thanks Gerry Share for helpful comments on this manuscript. I also thank two unnamed referees for especially helpful comments.